\documentclass[12pt,epsf,a4paper]{article}
\usepackage[english]{babel}
\usepackage{latexsym,epsfig}
\usepackage{amsmath}

0

\begin{document}
\baselineskip 6,5mm

\def\II{\relax{\rm 1\kern-.35em1}}
\def\IP{\relax{\rm I\kern-.18em P}}
\renewcommand{\theequation}{\thesection.\arabic{equation}}
\csname @addtoreset\endcsname{equation}{section}

\begin{flushright}
hep-th/9912022 \\
IC/99/182
\end{flushright}

\begin{center}

{}~\vfill

{ {\LARGE {Calibrated Geometries and Non Perturbative 
Superpotentials in M-Theory }}}

\vspace{20 mm}

{\bf {\large {Rafael Hern\'{a}ndez}}}$^{\dag}$

\vspace{8 mm}

{\em The Abdus Salam International Center for Theoretical Physics \\
Strada Costiera, 11. $34014$ Trieste, Italy}
\vspace{16 mm}

\end{center}


\begin{center}
{\bf Abstract}
\end{center}
  
\vspace{2 mm}
  
We consider non perturbative effects in M-theory compactifications on a 
seven-manifold of $G_2$ holonomy arising 
from membranes wrapped on supersymmetric 
three-cycles. When membranes are wrapped on 
associative submanifolds they induce 
a superpotential that can be calculated using calibrated geometry. This 
superpotential is also derived from compactification on a seven-manifold, 
to four dimensional Anti-de Sitter spacetime, 
of eleven dimensional supergravity 
with non vanishing expectation value of the four-form field strength.
  
\vspace{24 mm}

{\footnotesize \dag}\hspace{1 mm}{\footnotesize{\ttfamily e-mail address: rafa@ictp.trieste.it}}

\newpage


\section{Introduction}

The theory of minimal surfaces has been for a long time an area of active 
research in mathematics. A deep novelty on it was the theory of calibrations, 
introduced by Harvey and Lawson \cite{HL} in $1982$. A calibration is a closed 
$k$-form $\psi$ on a Riemannian manifold $M$ such that its restriction to each 
tangent plane of $M$ is less or equal to the volume of the plane. Submanifolds 
for which equality holds are said to be calibrated by $\psi$, and they have least 
volume in their homology class. 
With this defining property, minimal $p$-surfaces, or calibrated submanifolds, can be 
naturally employed to construct supersymmetric configurations of Dirichlet $p$branes in 
string theory, as in references \cite{BBS}-\cite{Marino}, an approach which is also 
probably able to shed some light \cite{SYZ,Vafa} on the study of the moduli space of 
calibrated submanifolds \cite{McLean,Hitchin}, whose global structure seems 
difficult to understand \cite{Gopakumar,JoyceSYZ}.
  
Motivated by reference \cite{Gukov}, where BPS solitons in the effective field theory 
arising from compactification of type IIA string theory, with non vanishing 
Ramond-Ramond fluxes, on a Calabi-Yau fourfold, were identified by Gukov with 
D-branes wrapped over calibrated submanifolds in the internal Calabi-Yau space to 
provide a simple and geometrical derivation of the superpotential in the two 
dimensional field theory \cite{Lerche,GVW}, in this paper we will consider 
compactification of M-theory on a seven-manifold of $G_2$ holonomy, which leads to  
a four dimensional field theory with $N=1$ supersymmetry, and apply ideas, as in 
\cite{Gukov}, of the theory of calibrations to understand the superpotential.
  
Superpotentials induced by membrane instantons in M-theory on a seven-manifold 
have been previously considered from the point of view of geometric engineering of 
field theories by Acharya in \cite{Acharya-Joyce}, where fractional membrane instantons, 
arising from compactification on Joyce orbifolds, are argued to generate a 
superpotential and, more recently, by Harvey and 
Moore in \cite{Harvey-Moore}, where membranes 
wrapped on rigid supersymmetric three-cycles are shown to induce non zero 
corrections to the superpotential, that can be expressed in terms of topological 
invariants 
of the three-cycle. In this paper we will argue how the allowed calibrations in $G_2$ 
holonomy seven-manifolds imply that the only contributions to the superpotential 
in the four dimensional field theory will come from membranes whose worldvolume has 
been wrapped on supersymmetric three-cycles of the internal manifold, and exclude 
posible corrections from eleven dimensional fivebranes, in the same way as in 
\cite{Gukov} various D-branes in type IIA string theory are shown to contribute 
to the superpotential in the two dimensional $N=(2,2)$ field theory, when 
wrapping calibrated submanifolds in the Calabi-Yau fourfold. However, we will not 
consider compactifications to four dimensional Minkowski spacetime, as in 
\cite{Harvey-Moore}, but to four dimensional Anti-de Sitter space, to avoid the 
constraint that the fourth rank tensor field strength must be vanishing \cite{Candelas}. 
This will allow a nice geometrical picture for the superpotential.
  
Relevant notions on calibrated geometries, and its realizations in supersymmetric 
compactifications, are described in section $2$. Section $3$ is devoted to the 
construction of the superpotential induced by membrane instantons, identifying instantons 
in the four dimensional field theory with membranes whose worldvolume is wrapped over 
associative submanifolds in the $G_2$ holonomy seven-manifold. The expression 
for the superpotential is justified in section $4$ by compactification of eleven 
dimensional supergravity, with non trivial fourth rank tensor field strength, on a 
manifold with $G_2$ holonomy to a four dimensional Anti-de Sitter spacetime. In section 
$5$ we present some concluding remarks and possible implications of the work presented 
here.


\section{Supersymmetry and the Calibration Bound}

In this section we will review some generalities about calibrated geometry\footnote{
For a more detailed and formal treatment on calibrated geometry we refer the 
reader to the original reference \cite{HL}, or \cite{Harvey} for a collection 
of results.}, and its 
relation to supersymmetric compactifications, and introduce some results and 
definitions relevant to the rest of the paper.
  
Given a Riemannian manifold $M$, with metric $g$, an oriented tangent $k$-plane 
$V$ on $M$ is a vector subspace $V$ of the tangent space $T_p M$ to $M$, with 
$\hbox {dim }(V)=k$, equipped with an orientation. The restriction $g|_V$ is the 
Euclidean metric on $V$, and allows to define, together with the orientation on 
$V$, a natural volume form $\hbox {Vol}_V$ on the tangent space $V$.
  
A closed $k$-form $\psi$ is said to be a {\em calibration on $M$} if for every 
oriented $k$-plane $V$ on $M$ it is satisfied
\begin{equation}
\psi|_V \leq \hbox{Vol}_V
\label{1}
\end{equation}
where, by $\psi|_V$, we mean
\begin{equation}
\psi|_V = \alpha \, . \, \hbox{Vol}_V 
\label{2}
\end{equation}
for some $\alpha \in {\bf R}$, so that condition (\ref{1}) holds only if $\alpha 
\leq 1$.
  
Now let $N$ be an oriented submanifold of $M$, with dimension $k$, so that each 
tangent space $T_p N$, for $p \in N$, is an oriented tangent $k$-plane. The 
submanifold $N$ is called a {\em calibrated submanifold} with respect to the calibration 
$\psi$ if
\begin{equation}
\psi|_{T_p N} = \hbox {Vol}_{T_p N}
\label{3}
\end{equation}
for all $p \in N$. It is easy to show that calibrated manifolds have minimal area in 
their homology class \cite{HL}; to see this, let us denote by $[ \psi ]$ 
the de Rham cohomology class, $[ \psi ] \in H^k(M;{\bf R})$, and by $[ N ]$ the 
homology class, $[ N ] \in H_k(M;{\bf R})$. Then,
\begin{equation}
[ \psi ] \, . \, [ N ] = \int _{p \in N} \psi|_{T_p N}.
\label{4}
\end{equation}
But condition (\ref{1}) implies
\begin{equation}
\int_{p \in N} \psi|_{T_p N} \leq \int_{p \in N} \hbox{Vol}_{T_p N} = 
\hbox{Vol}(N),
\label{5}
\end{equation}
so that $\hbox{Vol}(N) \geq [ \psi ] \, . \, [ N ]$. Equality holds 
for calibrated submanifolds \cite{HL}.
  
In this paper we will be interested in calibrated geometries in seven-dimensional 
Joyce manifolds with $G_2$ holonomy so that $M$, in what follows, will be a compact 
seven-manifold, $M = X_7$, with a torsion free structure, that we will denote 
by $\Psi^{(3)}$ (the holonomy group of the metric associated to $\Psi^{(3)}$ is 
$G_2$ if and only if $\pi_1(X_7)$ is finite \cite{Joyce}).
  
The covariantly constant three-form $\Psi^{(3)}$ constitutes the {\em associative 
calibration}, with respect to the set of all associative three-planes, which are 
the canonically oriented imaginary part of any quaternion subalgebra of ${\bf O}$. 
Calibrated submanifolds with respect to $\Psi^{(3)}$ will, in what follows, be 
denoted by $S_{\Psi}$, and referred to as {\em associative submanifolds}. The 
Hodge dual to the associative calibration is a four-form $^*\Psi^{(4)}$, known as 
the {\em coassociative calibration}, responsible for {\em coassociative 
submanifolds}.
  
The existence of associative submanifolds, volume minimizing, 
can be directly understood from the point of view of 
compactifications of M-theory on manifolds with $G_2$ holonomy \cite{BBM}. Let us 
see how this comes about. The low energy limit of M-theory is expected to be eleven 
dimensional supergravity, which is a theory that contains membrane solutions, described by the action \cite{BST,BST2} \footnote{We will chose units such that the eleven 
dimensional Planck length is set to one.}
\[
S_{\hbox{\tiny{Membrane}}} = \int d^3 \sigma \sqrt{h} [ \frac {1}{2} h^{\alpha \beta} 
\partial_{\alpha} X^M \partial_{\beta} X^N g_{MN} - \frac {1}{2} -i \bar{\theta} 
\Gamma^{\alpha} \nabla_{\alpha} \theta 
\]
\begin{equation}
+ \frac {i}{3 \, !} \epsilon^{\alpha \beta \gamma} C_{MNP} \partial_{\alpha} X^M 
\partial_{\beta} X^N \partial_{\gamma} X^P + \cdots ],
\label{6}
\end{equation}
where $X^M(\sigma^{i})$ represents the membrane configuration, $\theta$ is an eleven 
dimensional Dirac spinor and $h_{\alpha \beta}$ is the induced metric on the worldvolume 
of the membrane (upper case latin indices are defined in eleven dimensions, while greek 
indices $\alpha,\beta,\gamma$ label coordinates on the worldvolume).
  
On the membrane fields, global fermionic symmetries act as
\begin{eqnarray}
\delta_{\eta} \theta & = & \eta, \nonumber \\
\delta_{\eta} X^M    & = & i \bar{\eta} \Gamma^M \theta,
\label{7}
\end{eqnarray}
where $\eta$ is an eleven dimensional constant anticonmuting spinor. But the theory is 
also invariant under local fermionic transformations,
\begin{eqnarray}
\delta_{\kappa} \theta & = & 2 P_+ \kappa(\sigma), \nonumber \\
\delta_{\kappa} X^M    & = & 2i \bar{\theta} \Gamma^M P_+ \kappa (\sigma),
\label{8}
\end{eqnarray}
with $\kappa$ some eleven dimensional spinor, and $P_+$ a projection operator 
\cite{HLP}; the operators $P_{\pm}$ are defined as
\begin{equation}
P_{\pm} = \frac {1}{2} ( 1 \pm \frac {i}{3 \, !} \epsilon^{\alpha \beta \gamma}
\partial_{\alpha} X^M \partial_{\beta} X^N \partial_{\gamma} X^P \Gamma_{MNP}),
\label{9}
\end{equation}
and can be easily shown to satisfy $P_{\pm}^2=P_{\pm}$, $P_+ P_- =0$ and $P_+ 
+ P_- =1$.
  
As bosonic membrane configurations break all the global supersymmetries generated by 
$\eta$, unbroken supersymmetry remains only if there is a spinor $\kappa 
(\sigma)$ such that
\begin{equation}
\delta_{\kappa} \theta + \delta_{\eta} \theta = 2 P_+ \kappa (\sigma) 
+ \eta =0
\label{10}
\end{equation}
or, equivalently,
\begin{equation}
P_- [ 2 P_+ \kappa (\sigma) + \eta ] = P_- \eta =0.
\label{11}
\end{equation}
  
If the eleven dimensional theory is compactified on a seven-manifold of 
$G_2$ holonomy, finding a covariantly constant spinor $\eta$ satisfying 
condition (\ref{11}) is equivalent to the existence of a supersymmetric three-cycle 
in the seven-manifold, which is precisely the associative submanifold \cite{BBM}. 
To prove this, we will formally write the covariantly constant three-form 
(the associative calibration) as
\begin{equation}
\Psi^{(3)} = \frac {1}{3 \, !} \psi_{mnp} dX^m \wedge dX^n \wedge dX^p,
\label{12}
\end{equation}
and decompose the eleven dimensional spinor $\eta$ in terms of a four dimensional 
spinor, $\epsilon$, and the covariantly constant spinor $\xi$ on the manifold with 
$G_2$ holonomy (this spinor is unique up to scale),
\begin{equation}
\eta = \epsilon \otimes \xi.
\label{13}
\end{equation}
Now, if we chose a normalization such that the action of the seven dimensional 
gamma matrices on the covariantly constant spinor of the internal manifold is
\begin{equation}
\gamma_{mnp} \, \xi = \psi_{mnp} \, \xi,
\label{14}
\end{equation}
the condition (\ref{11}) for unbroken supersymmetry becomes
\begin{equation}
\frac {1}{2} \left( 1 - \frac {i}{3 \, !} \epsilon^{\alpha \beta \gamma} 
\partial_{\alpha} X^m \partial_{\beta} X^n \partial_{\gamma} X^p \psi_{mnp} 
\right) \xi =0,
\label{15}
\end{equation}
which is equivalent to \cite{BBM},
\begin{equation}
\frac {1}{3 \, !} \partial_{[ \alpha} X^m \partial_{\beta} X^n \partial_{\gamma ]} 
X^p \psi_{mnp} = \epsilon_{\alpha \beta \gamma},
\label{16}
\end{equation}
so that the pull back of the three-form is proportional to the volume element
\footnote{An identical condition can be similarly derived for the coassociative 
calibration \cite{BBM},
\[
\frac {1}{4 \, !} \partial_{[ \alpha} X^m \partial_{\beta} X^n \partial_{\gamma } 
X^p \partial_{\delta ]} X^q \psi_{mnpq} = \epsilon_{\alpha \beta \gamma \delta}.
\]}.
  
Condition (\ref{16}) implies that the membrane worldvolume, wrapped on the three-cycle 
calibrated by $\Psi^{(3)}$, has been minimized, as can be easily seen from the 
inequality
\begin{equation}
\int d^3 \sigma \sqrt{h} (P_- \xi)^{\dagger} (P_- \xi) \geq 0,
\label{17}
\end{equation}
as in \cite{BBS}. Using the projector (\ref{9}), and the fact that $P_-^{\dagger} 
P_- = P_-$, (\ref{17}) becomes
\begin{equation}
\int d^3 \sigma \sqrt{h} \, \bar{\xi} \xi \geq \int d^3 \sigma \, \bar{\xi} 
\frac {i}{3 \, !} \epsilon^{\alpha \beta \gamma} \partial_{\alpha} X^m 
\partial_{\beta} X^n \partial_{\gamma} X^p \psi_{mnp} \xi,
\label{18}
\end{equation}
or $V_3 \geq \int_{S_{\Psi}} \Psi^{(3)}$. The bound is saturated if and only if 
$P_- \xi=0$, which is precisely the condition for unbroken supersymmetry.


\section{Associative Calibrations and Superpotential for M-Theory Compactifications}

M-theory compactification on a seven-manifold with $G_2$ holonomy produces 
a four dimensional field theory with $N=1$ supersymmetry. At low energy, trusting 
eleven dimensional supergravity as an aproximation to M-theory, the effective four 
dimensional supergravity theory describing the massless modes is $N=1$ 
supergravity coupled to $b_2$ vector multiplets, and $b_3$ chiral multiplets, 
where $b_2$ and $b_3$ are Betti numbers of $X_7$ \cite{Papado}. However, in this 
paper we will not be interested in this relatively poor, from the point of 
view of physics, spectrum, or posible increases of interest when singularities are allowed 
in the seven-manifold, as in \cite{Acharya-Joyce}. What we will wonder about is 
the generation of a non perturbative superpotential, arising in the effective four 
dimensional field theory from M-theory effects, applying the ideas in \cite{Gukov}.
  
In reference \cite{Gukov}, the generation of a 
superpotential in the two dimensional theories obtained when compactifying type IIA 
string theory on Calabi-Yau fourfold with background Ramond-Ramond fluxes was 
considered, identifying BPS solitons in the field theory with D-branes wrapped 
over calibrated submanifolds in the internal manifold. In this section, following the 
same reasoning, we will wrap the M-theory membrane worldvolume over some associative 
submanifold in the seven-manifold $X_7$, $S_{\Psi} \in H_3(X_7, {\bf Z})$, which is 
a supersymmetric three-cycle, defined through condition (\ref{16}). This state, 
from the point of view of the four dimensional field theory, is an instanton. The 
flux of the fourth-rank antisymmetric field strength, $F=dC$, associated to the 
membrane, over the ``complement'' $(S_{\Psi})^{\perp}$ in $X_7$, which will be 
some four dimensional manifold, $Y_4$, jumps by one when crossing the membrane. 
To understand this, two points $P$ and $Q$ can be chosen to ly at each side of the 
three-cycle $S_{\Psi}$, with the membrane worldvolume wrapped on it. If $T$ is a 
path joining these two points, and it is chosen to intersect the membrane at a single 
point, the five-manifold $Y_4 \times T$ will also intersect the membrane at a single 
point. Now, as $\frac {dF}{2 \pi}$ is a delta function for the membrane,
\begin{equation}
\int_{Y_4 \times T} \frac {dF}{2 \pi} =1
\label{a}
\end{equation}
or, once Stoke's theorem is employed,
\begin{equation}
\int_{Y_4 \times P} \frac {F}{2 \pi} - \int_{Y_4 \times Q} \frac {F}{2 \pi} =1.
\label{b}
\end{equation}
Hence, the field theory vacua connected by the instanton solution will correspond 
to different four-form fluxes, $F_1$ and $F_2$; the variation $\frac {\Delta F}
{2 \pi}$ is then Poincar\'e dual to the homology class of the supersymmetric three-cycle, 
$[S_{\Psi}]$. But the amplitude for the configuration of the membrane is 
proportional to the volume of the associative submanifold, $S_{\Psi}$,
\begin{equation}
\int_{S_{\Psi}} \Psi^{(3)},
\label{19}
\end{equation}
the constant of proportionality being simply the membrane tension. If we denote 
the superpotential in the field theory by $W$, the amplitude of the instanton 
connecting two vacua will be given by the absolute value of $\Delta W$; then, when 
(\ref{b}) is taken into account, (\ref{19}) becomes
\begin{equation}
\Delta W = \int_{S_{\Psi}} \Psi^{(3)} = \frac {1}{2 \pi} \int_{X_7 =
S_{\Psi} \times Y_4} \Psi^{(3)} \wedge \Delta F^{(4)},
\label{20}
\end{equation}
so that we expect a superpotential
\begin{equation}
W = \frac {1}{2 \pi} \int_{X_7} \Psi^{(3)} \wedge F^{(4)},
\label{21}
\end{equation}
for compactification of M-theory on a seven-manifold of $G_2$ holonomy. Expression 
(\ref{21}) is the analogous of the untwisted and twisted chiral superpotentials 
obtained in \cite{Gukov} for compactification of type IIA string theory in a 
Calabi-Yau fourfold with non-vanishing Ramond-Ramond fluxes.
  
We could now also wonder about the possible contribution of M-theory fivebranes 
to the superpotential. In fact, the result that $\pi_1(X_7)$ is finite if and only if 
the holonomy is in $G_2$ \cite{Joyce}, implies that $H_6(X_7;{\bf Z})$ is also 
finite, so that we might expect instanton effects coming from the fivebrane. However, 
there is no six dimensional calibrated submanifold in $X_7$, because the only 
calibrations are $\Psi^{(3)}$ and $^*\Psi^{(4)}$, so that there will be no BPS 
fivebrane instanton contribution to the superpotential.


\section{Eleven Dimensional Supergravity \\on Seven-Manifolds}

In this section we will present evidence for the superpotential (\ref{21}) from 
supersymmetric compactification of M-theory on $AdS_4 \times X_7$. Compactification 
of eleven dimensional supergravity on a seven-manifold is very restrictive, because the 
four dimensional non compact spacetime is a supersymmetric Minkowski space only 
if all components of the fourth rank antisymmetric field strength vanish, $F=0$, and 
the internal manifold is Ricci-flat \cite{Candelas}. However, if some of these 
conditions are relaxed, more general compactifications can be performed. In 
\cite{Gunaydin,deWit}, compactification of eleven dimensional supergravity to four 
dimensional Anti-de Sitter spacetime was shown to allow a four-form 
field strength proportional to the cosmological constant of the external space, and 
which therefore vanishes for compactifications to Minkowski spacetime. The requirements 
on the four-form $F$, assuming that supersymmetry remains unbroken, with zero 
cosmological constant, for compactification of M-theory with a warp factor 
on a Calabi-Yau fourfold have been obtained in 
\cite{BB}, and extended to a three dimensional Anti-de Sitter external space in \cite{GVW}. 
The analysis of this section, where compactification of M-theory to a four dimensional 
spacetime, with non vanishing cosmological constant, on an internal manifold with $G_2$ 
holonomy will be considered, will hence follow closely those in references \cite{BB,GVW,Gukov}.
  
The bosonic form of the eleven dimensional effective action looks like 
\begin{equation}
S= \frac {1}{2} \int d^{11} x \sqrt{g} [ R - \frac {1}{2} F^{(4)} \wedge 
* F^{(4)} - \frac {1}{6} C^{(3)} \wedge F^{(4)} \wedge F^{(4)} - C^{(3)} 
\wedge I^{(8)}],
\label{22}
\end{equation}
where the gravitational Chern-Simons correction, associated to the sigma-model 
anomaly of the six dimensional fivebrane worldvolume \cite{Duff}, can be expressed in terms 
of the Riemann tensor \cite{Alvarez-Gaume},
\begin{equation}
I_8 = - \frac {1}{768} (\hbox {tr }R^2)^2 + \frac {1}{192} \hbox {tr }R^4.
\label{23}
\end{equation}
The complete action is invariant under local supersymmetry transformations \cite{CJS}
\begin{eqnarray}
\delta_{\eta} e^{A}_M & = & i \bar{\eta} \Gamma^{A} \psi_M, \nonumber \\
\delta_{\eta} C_{MNP} & = & 3 i \bar{\eta} \Gamma_{[MN} \psi_{P]}, \nonumber \\
\delta_{\eta} \psi_M  & = & \nabla_M \eta - \frac {1}{288} (\Gamma_M^{PQRS} -8 
\delta_M^{P} \Gamma^{QRS}) F_{PQRS} \eta,
\label{24}
\end{eqnarray}
where $e^{A}_M$ is the elfbein, $\psi_M$ the gravitino, $\eta$ an eleven dimensional 
anticonmuting Majorana spinor, and $\nabla_M$ the covariant derivative, involving the 
Christoffel connection.
  
A supersymmetric configuration exists if and only if the above transformations vanish 
for some spinor $\eta$. If the background is Lorentz covariant in the 
four dimensional spacetime, the 
background spinor $\psi_M$ must vanish, so that $e^{A}_M$ and $C_{MNP}$ are unchanged 
by the supersymmetry transformation. Hence, the only constraint that remains to be 
impossed is
\begin{equation} 
\nabla_M \eta - \frac {1}{288} (\Gamma_M^{PQRS} -8 
\delta_M^{P} \Gamma^{QRS}) F_{PQRS} \eta =0.
\label{26}
\end{equation}
  
The most general metric, maximally symmetric, for compactification on $X_7$,
\begin{equation}
ds_{11}^2 = \Delta(x^m)^{-1} (ds_4^2(x^{\mu}) + ds_7^2(x^m)),
\label{27}
\end{equation}
includes a scalar function called the warp factor, $\Delta(x^m)$, depending on the 
internal dimensions. We are now choosing a notation 
such that the eleven dimensional upper case indices split into four dimensional 
greek indices, while latin indices label the set of 
coordinates tangent to the internal manifold. 
With the choice (\ref{27}) for the metric, condition (\ref{26}) becomes \cite{BB}, 
\begin{equation} 
\nabla_M \eta - \frac {1}{4} \Gamma_M^N \partial_N (\log \Delta) \eta - 
\frac {\Delta^{3/2}}{288} (\Gamma_M^{PQRS} -8 
\delta_M^{P} \Gamma^{QRS}) F_{PQRS} \eta =0.
\label{30}
\end{equation}
  
Now, let us decompose the gamma matrices in the convenient $11=7+4$ split,
\begin{eqnarray}
\Gamma^{\mu} & = & \gamma^{\mu} \otimes \II, \nonumber \\
\Gamma^{m}   & = & \gamma^5 \otimes \gamma^m,
\label{31}
\end{eqnarray}
where $\gamma^5 = \frac {i}{4 \, !} \epsilon_{\mu \nu \rho \sigma} \gamma^{\mu} \gamma^{\nu}
\gamma^{\rho} \gamma^{\sigma}$ is the four dimensional chirality operator,
so that $(\gamma^5)^2 = + \II$, and $\gamma^{m}$ are chosen such that 
$\frac {i}{7 \, !} g^{1/2} \epsilon_{m_1 \ldots m_7} \gamma^{m_1} \ldots \gamma^{m7} = 
+ \II.$
We will also choose, as in \cite{deWit}, the most general covariant form for the fourth rank 
antisymmetric tensor field strength\footnote{In compactifications of string theory 
or M-theory on a Calabi-Yau fourfold a global anomaly, given by the Euler number 
of the Calabi-Yau fourfold, arises \cite{SVW,W}. This anomaly can be cancelled 
if non zero background fluxes are allowed, and/or strings or membranes are introduced 
filling, respectively, two or three dimensional spacetime, for compactification of 
string theory or M-theory. These strings, or membranes, are represented by maximally 
symmetric tensor fields. However, for seven-manifolds there is no such interpretation 
for the ansatz of the first equation in (\ref{34}).} 
\begin{eqnarray}
F_{\mu \nu \rho \sigma} & = & m \epsilon_{\mu \nu \rho \sigma}, \nonumber \\
F_{\mu \nu \rho s}      & = & F_{\mu \nu r s} = F_{\mu n r s} =0,
\label{34} 
\end{eqnarray}
while
\begin{equation}
F_{mnpq} \hbox { arbitrary},
\label{34a}
\end{equation}
where $m$ can depend upon the extra (internal) dimensions, 
and a decomposition for the supersymmetry parameter, $\eta = \epsilon 
\otimes \zeta$. 
  
With the split (\ref{31}), and the above ansatz for $F$, the $\mu$-component of the 
supersymmetry condition (\ref{30}) becomes
\[
\nabla_{\mu} \eta - \frac {1}{4} \gamma_{\mu} \gamma^5 \otimes \gamma^{n} \partial_n 
(\log \Delta) \eta - \frac {\Delta^{3/2}}{288} \gamma_{\mu} \gamma^{mnpq} F_{mnpq} \eta \,
\]
\begin{equation}
+ \frac {1}{6} \Delta^{3/2} i m \gamma_5 \gamma_{\mu} \eta = 0.
\label{35}
\end{equation}
When $\epsilon$ is a four dimensional anti-commuting Killing spinor satisfying 
$ \nabla_{\mu} \epsilon = \frac {\Lambda}{2} \gamma_{\mu} \epsilon $, 
equation (\ref{35}), with the decomposition $\eta = \epsilon 
\otimes \zeta$, leads to the solution
\begin{eqnarray}
144 \Lambda \zeta & = & \Delta^{3/2} F_{mnpq} \gamma^{mnpq} \, \zeta, \nonumber \\
m \, \zeta & = & i \gamma^n \partial_n \Delta^{-3/2} \zeta.
\label{38}
\end{eqnarray}
  
Similarly, with the decomposition (\ref{31}), the $m$-component of condition 
(\ref{30}) becomes, after some gamma 
matrices algebra\footnote{A complete collection of commutators and anticommutators, 
and useful gamma matrices identities, can be found in the appendices in 
\cite{Candelas} and \cite{Candelas2}, where the $11=7+4$ split is also considered.},
\[
\nabla_m \eta - \frac {\Lambda}{2} \gamma_m \eta + \frac {1}{4} \partial_m 
(\log \Delta) \eta - \frac {3}{8} \partial_n (\log \Delta) \gamma_m^n 
\eta 
\]
\begin{equation}
+ \frac {\Delta^{3/2}}{24} F_{mpqr} \gamma_5 \gamma^{pqr} 
\eta =0,
\label{40}
\end{equation}
where we have made use of (\ref{38}). As is 
\cite{BB,GVW}, the transformed quantities 
\begin{eqnarray}
\tilde{g}_{mn} & = & \Delta^{-3/2} g_{mn}, \nonumber \\
          \rho & = & \Delta^{1/4} \eta,
\label{41}
\end{eqnarray}
lead equation (\ref{40}) to the simpler form
\begin{equation}
\tilde{\nabla}_m \rho - \frac {\Lambda}{2}\Delta^{3/4} \tilde{\gamma}_m \rho 
+ \frac {1}{24} \Delta^{-15/4} F_{mpqr} \tilde{\gamma}_5 \tilde{\gamma}^{pqr} \rho =0.
\label{42}
\end{equation}
Now, if we decompose the eleven dimensional spinor $\rho$ as $\xi \otimes \varepsilon$, 
with $\xi$ chosen to be the covariantly constant spinor on the seven-manifold of 
$G_2$ holonomy, and we use the fact that the associative calibration $\Psi^{(3)}$ satisfies 
relation (\ref{14}), $\gamma^{pqr} \xi = \psi^{pqr} \xi$, then the components of 
(\ref{42}) reproduce, 
when integrated over the volume of $X_7$, the form of the superpotential proposed 
in (\ref{21}), with 
a coefficient related to the warp factor, depending hence on the internal dimensions, 
as a consequence of the ansatz (\ref{34}), and with the cosmological constant 
$\Lambda$ identified with the vacuum value of the superpotential, $W$, in the same 
way as the mass, and the twisted mass, are identified in \cite{Gukov} with the 
vacuum values of $W$ and $\tilde{W}$. Note however the appearance of the chirality operator 
in expression (\ref{42}).


\section{Conclusions}

In this paper we have investigated the generation of a non perturbative 
superpotential in the four dimensional field theories obtained from compactification 
of M-theory on a seven-manifold with $G_2$ holonomy. The allowed calibrations 
in these manifolds imply that the only BPS instantons are those obtained 
wrapping the worldvolume of the M-theory membrane over associative submanifolds, which are 
the analogous of special Lagrangian cycles in Calabi-Yau manifolds, while 
instantons coming from fivebranes wrapping six-cycles in the seven-manifold are 
not BPS states. An analysis relying on calibrated submanifolds can also be used to 
study compactification of M-theory on a Calabi-Yau fourfold, where fivebranes 
wrapped over (complex) codimension one divisors $D$ are 
however known to contribute to the superpotential in the three dimensional 
field theory arising upon compactification of M-theory on the fourfold 
if the divisor satisfies the topological requirement that $\chi({\cal O}_D,D)=1$ 
\cite{Wnpsp}. This precise condition can probably be related to the fact that fourfolds 
with $SU(4)$ holonomy admit, besides from Cayley calibrations, what are known 
as K\"ahler calibrations, which are obtained when considering powers of the complexified 
K\"ahler form,
\begin{equation}
\Psi = \frac {1}{p \, !} {\cal K}^p,
\label{43}
\end{equation}
because the fact that ${\cal K}$ is covariantly constant ensures that $\Psi$, as 
defined in (\ref{41}), is 
also covariantly constant. Submanifolds calibrated by $\Psi$ are complex 
submanifolds, of dimension $p$, so that fivebranes will become BPS 
instantons once they are wrapped over some (complex) three dimensional manifold, 
calibrated by $\frac {1}{3 \, !} {\cal K}^3$. 
  
In reference \cite{JoyceSYZ}, singularities of special Lagrangian three-cycles, 
and their compactness properties, were studied in detail, and a topological 
invariant, counting special Lagrangian homology three-spheres, was proposed. 
If an equivalent analysis can be repeated on seven-manifolds with $G_2$ 
holonomy, then an analogous topological invariant can probably be constructed, counting 
associative submanifolds or, which is more interesting physically, counting the 
number of membranes, or instantons, in the 
homology class $[S_{\Psi}]$, an idea closely related 
to the Gromov-Witten invariants, counting pseudo-holomorphic curves in 
symplectic manifolds. We hope to address this study in a subsequent paper.

\vspace{8 mm}

{\bf Acknowledgements}

It is a pleasure to thank E. C\'aceres, A. Mukherjee, K. Narain and K. Ray for 
useful discussions, and T. Ali and G. Thompson for comments on the manuscript. This research is partly supported by the EC contract no.
ERBFMRX-CT96-0090.

\newpage


\end{document}